\documentclass[sigconf]{acmart}

\usepackage{caption}
\setkeys{Gin}{width=\linewidth,totalheight=\textheight,keepaspectratio}
\makeatletter
\AtBeginDocument{%
  \geometry{twoside=true, paperwidth=8.5in, paperheight=11in,%
    includeheadfoot=false, columnsep=2pc,%
    top=1in, bottom=1in, inner=0.75in, outer=0.75in,%
    marginparwidth=2pc, heightrounded}%
  \fancypagestyle{standardpagestyle}{%
    \fancyhf{}%
  }%
  \fancypagestyle{firstpagestyle}{%
    \fancyhf{}%
  }%
}
\makeatother

\usepackage[ruled,vlined,linesnumbered]{algorithm2e}
\SetAlFnt{\footnotesize}

\usepackage{listings} %

\definecolor{codegray}{rgb}{0.5,0.5,0.5}   %
\definecolor{codegreen}{rgb}{0,0.6,0}      %
\definecolor{codepurple}{rgb}{0.58,0,0.82} %
\definecolor{codered}{rgb}{0.64,0.08,0.08} %
\definecolor{codeblue}{rgb}{0.11,0.63,0.89} %

\definecolor{codegreen30}{rgb}{0.70, 0.88, 0.70} %
\definecolor{codepurple30}{rgb}{0.87, 0.70, 0.95} %
\definecolor{codered30}{rgb}{0.89, 0.72, 0.72}   %
\definecolor{codeblue30}{rgb}{0.73, 0.89, 0.97}  %

\lstdefinestyle{mystyle}{
    commentstyle=\color{codegreen},
    keywordstyle=\color{codered},
    numberstyle=\tiny\color{codegray},
    stringstyle=\color{codepurple},
    basicstyle=\ttfamily\scriptsize,
    breakatwhitespace=false,         
    breaklines=true,
    captionpos=t,
    keepspaces=true,
    numbers=left,
    numbersep=5pt,
    showspaces=false,
    showstringspaces=false,
    showtabs=false,
    tabsize=2,
    frame=single,
    xleftmargin=2em,
    framexleftmargin=1.5em
}
\lstdefinestyle{armstyle}{
    commentstyle=\color{codegreen},
    keywordstyle=\color{codered},
    numberstyle=\tiny\color{codegray},
    stringstyle=\color{codepurple},
    basicstyle=\ttfamily\tiny,
    breakatwhitespace=false,         
    breaklines=true,
    captionpos=t,
    keepspaces=true,
    numbers=left,
    numbersep=5pt,
    showspaces=false,
    showstringspaces=false,
    showtabs=false,
    tabsize=2,
    frame=single,
    xleftmargin=2em,
    framexleftmargin=1.5em,
    morekeywords={stp, ldp, adrp, ret, cbnz, cbz},
    ndkeywords={x0,x1,x2,x3,x4,x5,x6,x7,x8,x9,x10,x11,x12,x13,x14,x15,x16,x17,x18,x19,x20,x21,x22,x23,x24,x25,x26,x27,x28,x29,x30,sp,lr}
}

\usepackage{subcaption}

\newcommand{\colorlegend}[1]{{\color{#1}\rule{3ex}{1.5ex}}\hspace{1pt}}

\usepackage{booktabs}
\usepackage{multirow}

\usepackage{cleveref}

\AtBeginDocument{%
  }

\setcopyright{cc}
\setcctype{by}
\copyrightyear{2026}
\acmYear{2026}
\acmDOI{10.1145/3770743.3804170}
\acmConference[DAC '26]{63rd ACM/IEEE Design Automation Conference}{July 26--29,
  2026}{Long Beach, CA, USA}
\acmBooktitle{63rd ACM/IEEE Design Automation Conference (DAC '26),
  July 26--29, 2026, Long Beach, CA, USA}
\acmISBN{979-8-4007-2254-7/2026/07}

\begin{document}

\title{Janus: Compiler-Based Defense Against Transient Execution Attacks Using ARM Hardware Primitives}

\author{Ciyan Ouyang, Peinan Li*, Yubiao Huang, Dan Meng, and Rui Hou}
\affiliation[obeypunctuation=true]{%
  \institution{State Key Laboratory of Cyberspace Security Defense\\Institute of Information Engineering, CAS, Beijing, China}
  \city{\mbox{}}
  \country{\mbox{}}
}

\renewcommand{\shortauthors}{Ouyang et al.}

\begin{abstract}
We present \textbf{Janus}, a compiler-based security framework that mitigates transient execution attacks like Spectre and control-flow hijacking on ARM64 platforms. Janus integrates speculative execution and control flow dependencies with PA modifiers, using PA and BTI microarchitectural features to prevent control-flow speculation attacks and secure both control flow and speculative execution through existing control-flow integrity mechanisms. To optimize performance, Janus minimizes overhead by merging defense operations across different defense layers (modifier fusion) and reusing registers of protected variables (carrier reuse), while maintaining strong security guarantees. Evaluation on SPEC CPU2017 shows an average performance overhead of 3.85\%, with real-world applications exhibiting overheads ranging from 2.97\% to 7.80\%. Janus offers effective speculative execution security and low performance and code size overhead, making it a robust solution for ARM-based systems.

\end{abstract}



\maketitle

\begingroup
\renewcommand{\thefootnote}{}
\footnotetext{* Corresponding author: Peinan Li (lipeinan@iie.ac.cn)}
\endgroup

\pagestyle{empty}
\thispagestyle{empty}

\section{Introduction}
Speculative execution in modern processors has led to Spectre attacks\cite{kocher2020spectre}, which exploit microarchitectural side channels to leak isolated data, threatening system confidentiality. While several mitigations have been proposed, none fully balance security, performance, and deployability. Hardware-level solutions, such as modifying processor components or designing new instructions, have long deployment cycles and limited adaptability, offering inadequate protection for existing hardware and emerging threats. Industry and consumers urgently require defense solutions that can be quickly deployed on current hardware.

Prevailing software defenses fall into two categories: full speculation barriers, like lfence\cite{lfence}, which offer security but cause significant performance degradation, and partial mitigations, like SLH\cite{slh}, which reduce attack surfaces by identifying leakage gadgets. However, pattern-matching approaches struggle to cover diverse attack gadgets and often lose effectiveness with complex control-flow semantics.
Existing software defenses fail to address the security issues of speculative execution. Complete speculation barriers incur high costs, while gadget-based\cite{wiebing2024inspectre} defenses can't guarantee long-term security. 

A deployable defense on current hardware should: (1) abort illicit speculation without affecting legitimate speculation, (2) verify speculation trustworthiness at its source, and (3) constrain speculation targets to limit attackers' ability to create malicious speculation chains.
Fortunately, existing hardware primitives\cite{hadrwareprimitives}, such as Arm's Branch Target Identification (BTI) and Pointer Authentication (PA)\cite{armaspec}, can be repurposed to achieve these goals through their microarchitectural side effects, a concept we validate experimentally in \cref{sec:4:verify}.

BTI's property of squashing invalid speculative jumps enforces characteristic 3 by \textbf{restricting all speculative indirect branches} to valid BTI-guarded targets. The key to characteristics 1 and 2 lies in PA's microarchitectural side effect: a failed verification during speculation squashes the pipeline instead of raising an architectural fault. Janus operationalizes this by using a \texttt{csel} instruction to \textbf{conditionally poison a PA modifier} based on the true branch outcome. On a mis-speculated path, the poisoned PA \texttt{modifier} guarantees the verification will fail, thereby triggering a \textbf{targeted speculation squash} at the point of divergence and eliminating the need for gadget identification.

Janus's defense uses BTI to create a zero-overhead, coarse-grained security baseline for indirect branches, covering both architectural and microarchitectural states. This prevents jumps to illicit gadgets and allows PA to focus on fine-grained, cryptographically backed control-flow binding and speculative state verification, ensuring comprehensive security with minimal PA instruction overhead. In summary, the contributions are fourfold:

\begin{itemize}
    \item \textbf{Unified Framework for Control-Flow Speculation Defense:} We introduce the first framework leveraging memory safety hardware to provide Control-Flow Integrity (CFI)\cite{burow2017cfi} and control data integrity at both architectural and microarchitectural levels, defending against Spectre attacks V1\cite{kocher2020spectre}, V2\cite{bhi}, V5\cite{Koruyeh2018SpectreRS}) and mitigating CFI threats like ROP/JOP and DOP.

    \item \textbf{Signature Neutralization for Metadata Protection:} We propose a "signature neutralization" mechanism that uses \texttt{csel} to dynamically encode speculative state and metadata, invalidating signatures on mis-speculated paths to prevent leakage.

    \item \textbf{Optimized Synergistic Defense Implementation:} We optimize memory safety and speculative execution defense by reusing existing BTI and PA instructions, reducing instrumentation and refining the set of variables requiring protection. By analyzing variables shared by both attack classes, we further refine and minimize the set of variables needing protection.

    \item \textbf{Comprehensive Evaluation on ARMv9 Hardware:} We evaluate Janus on \texttt{ARMv9} hardware, showing performance overhead below 5\% on SPEC CPU 2017, successful compilation of real-world applications like Nginx, and verified security using Spectre proof-of-concept (PoC) exploits and the publicly available PACMAN exploit.
\end{itemize}

\section{Background}

\subsection{Spectre Attacks}

Spectre attacks manipulate processor predictions to access isolated data and leak it through microarchitectural side channels. Spectre attacks fall into two categories: control-flow Spectre (e.g., V1, V2, V5) exploiting branch prediction errors, and data-flow Spectre (e.g., V4) exploiting memory dependency mispredictions. A more dangerous hybrid threat, Speculative Memory-error Abuse (SMA)\cite{christou2024eclipse}, combines speculative execution with memory corruption. These attacks exploit speculative execution's ability to suppress fatal exceptions, leaking secret metadata used by defenses like cryptographic Pointer Authentication Code\texttt{(PAC)}\cite{pacman} and ASLR\cite{hackblind}. Once these defenses are compromised, attackers can escalate to more reliable architectural attacks, forming a dangerous attack cycle. Existing software mitigations are ineffective against these advanced hybrid threats.

\subsection{Memory Safety and Hardware Protection Primitives}

Memory safety prevents programs from accessing incorrect memory locations, avoiding vulnerabilities like buffer overflows\cite{cowan2000bufferoverflow} and use-after-free errors\cite{memoryvulners}. These flaws can lead to data corruption, information leakage, control-flow hijacking\cite{roemer2012return,bletsch2011jump,schuster2015counterfeit} or no-control data attacks\cite{controlflowbending}. Control-flow hijacking attacks manipulate the execution flow of programs, while no-control data attacks target the data integrity of applications; CFI and Data flow integrity DFI\cite{castro2006dfi} are security techniques designed to prevent these types of vulnerabilities. Modern CPUs enforce security through hardware mechanisms. Janus builds on Arm PA for fine-grained pointer integrity using a \texttt{PAC}. The hardware verifies this signature before pointer use, invalidating it if verification fails. Arm BTI provides coarse-grained control-flow protection, ensuring indirect branches target only legal locations. Similar primitives exist on x64 and RISC-V, making Janus cross-platform compatible.

\subsection{Threat Model}

We assume the platform enforces a Write Xor Execute \texttt{(W$\oplus$X)}\cite{Schink2019TakingAL,Pomonis2017kRXCK} policy, preventing code injection attacks, and that no other memory safety or Spectre defenses are in place. The attacker aims to compromise the confidentiality of critical secrets and exploit vulnerabilities in memory safety and control-flow Spectre gadgets. The attacker has the following capabilities:

\begin{itemize}
    \item An \textbf{arbitrary write primitive} to corrupt code pointers or control-flow data.
    \item The ability to \textbf{influence the speculative outcomes} of conditional branches and the targets of indirect branches in the victim program.
    \item The ability to monitor CPU microarchitectural state via a \textbf{side channel}, enabling the exfiltration of transiently accessed data. The attacker and victim may execute on the same physical machine.
\end{itemize}

We consider attacks that manipulate data-flow speculation to be orthogonal to the defenses provided by Janus.

\section{Motivation and Design Goals}

\begin{figure}[b]
        \centering
        \includegraphics[width=0.90\linewidth]{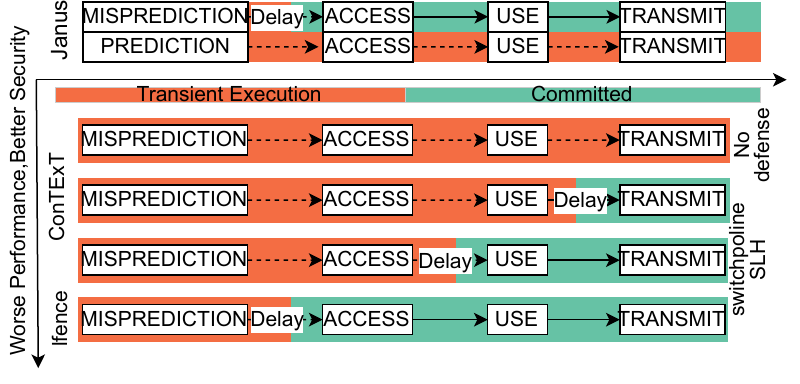}
        \vspace{-1.0em}
        \caption{Janus in the Hierarchy of Transient Execution Defenses}
        \label{fig:janus_hierarchy} %
        \vspace{-0.3em}
\end{figure}

Traditional memory safety mechanisms weren’t designed for Transient Execution Attacks (TEAs)\cite{Xiong2021SurveyOT,Fiolhais2023TransientExecutionAA}, leaving metadata used by defenses like ASLR\cite{team2003aslr} and stack canaries\cite{Tan2024IsTC} vulnerable. Attacks like PACMAN show how transient execution can compromise memory safety hardware, such as brute-forcing pointer authentication. Additionally, hardware-based defenses like FineIBT lack explicit protection against speculative execution. As shown in \cref{fig:janus_hierarchy}, TEA defenses typically operate at the backend of the attack pipeline, providing weaker security and failing to optimize synergistically with architectural memory safety. Janus combines architectural and microarchitectural security primitives, leveraging their side effects with intelligent compiler instrumentation. We chose the ARM platform for its mature ecosystem, but note that Janus’s core design is portable to x86 and RISC-V (similar hardware primitives\cite{intel2019cet,riscv2024cfi} exist or are coming soon).

The loss of CFI is a critical threat to memory safety and aligns closely with defending against transient execution attacks. Janus focuses on enforcing CFI, validating speculative control flow at the processor’s branch decision stage. This approach ensures synergistic protection at both architectural and microarchitectural levels.

\section{Verifying the Microarchitectural Foundations of Janus}
\label{sec:4:verify}

The design of Janus is predicated on the principle that modern hardware security features possess undocumented or under-utilized microarchitectural side-effects that can be repurposed to defend against transient execution attacks. While official ARM documentation and academic works like the PACMAN allude to these behaviors \cite{armaspec,pacman}, their reliability and specific mechanisms have not been deeply investigated for defensive use. This chapter provides the first empirical validation of two such critical properties on the ARMv8-A architecture, which form the foundation of our defense.

\begin{lstlisting}[
  float=htb,  %
  label={lst:pa_verification},
  language=C++, 
  caption={The Speculation-Squashing Effect of PA.}
]
void test_pa_speculation_squash(bool condition){
  // Trigger misprediction on `if`.
  MISTRAIN_BRANCH_PREDICTOR(condition);
  if (condition) {
    // Barrier: PAC validation fails here.
    VERIFY_PAC(CORRUPTED_POINTER);
    // Transmitter: Reached only if barrier fails.
    MEM_SIDECHANNEL_ACCESS(secret_A);
  }
}
\end{lstlisting}
\begin{lstlisting}[
  float=htb,  %
  label={lst:bti_verification},
  language=C++, 
  caption={The Speculative Redirection Deterrence of BTI.}
]
void NON_BTI_TARGET(){// Barrier: the absence of a BTI.
  // Transmitter: Reached only if barrier fails.
  MEM_SIDECHANNEL_ACCESS(secret_B);
}

void test_bti_speculation_deterrence(void (*fptr)()){
  // Mistrain BTB to point to the invalid target.
  MISTRAIN_BTB_TO(&NON_BTI_TARGET);
  // speculatively call NON_BTI_TARGET.
  fptr();
}
\end{lstlisting}

\begin{figure*}[thb]
    \centering %
    \captionsetup{skip=0pt}
    \includegraphics[width=0.96\textwidth]{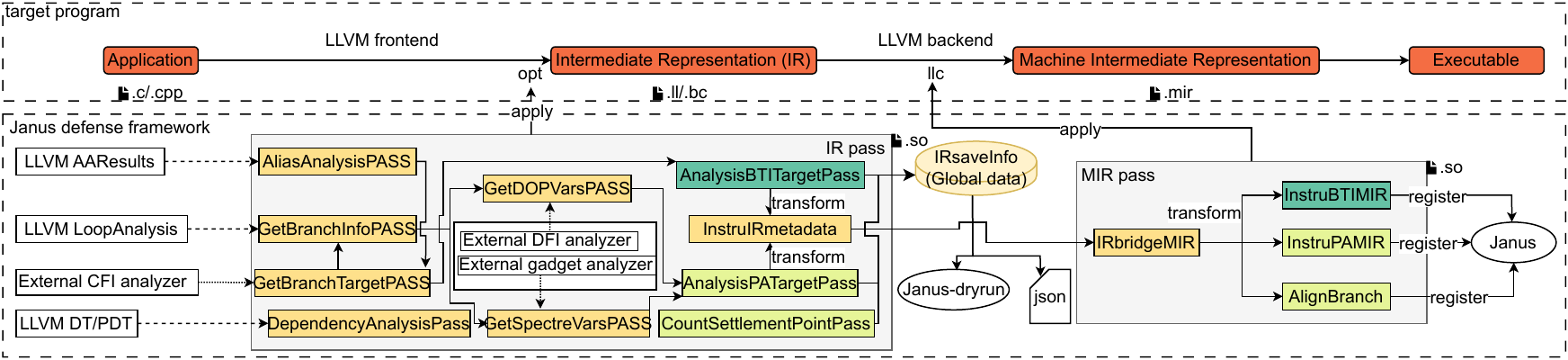}
    \caption{The architecture of the Janus compiler framework, detailing the pipeline of analysis, transformation, and instrumentation passes.}
    \label{fig:janus_implement}
    \vspace{-0.8em}
\end{figure*}

First, we investigate Arm PA, which architecturally faults on a \texttt{PAC} mismatch. We hypothesize that a \texttt{PAC} validation failure occurs early enough in the pipeline to prevent mis-speculated instructions from executing and leaking data(\cref{lst:pa_verification}). Second, we examine Arm BTI, which prevents control flow from landing at illegal targets. We hypothesize that a BTB mistrain to a non-BTI location will be detected, stalling or squashing the speculative pipeline(\cref{lst:bti_verification}).

To test these hypotheses, we conducted microbenchmark tests on the \texttt{Radxa Orion O6} platform, used in the evaluation in \cref{sec:7:evaluation}. This platform features \texttt{FEAT\_FPAC} and \texttt{FEAT\_FPACC\_SPEC} \cite{Arm_FEAT_FPAC_2023}, which are critical for pointer authentication and speculative execution control. \texttt{FEAT\_FPAC} allows the processor to verify pointer authenticity via cryptographic signatures, while \texttt{FEAT\_FPACC\_SPEC} ensures speculative control flow integrity. Using performance counters and a timing-based side channel, we confirmed that \texttt{PAC} validation failure and BTI target mismatches prevent transient execution of gadgets. Specifically, we observed that the cache did not read erroneous branch data during mis-speculation, confirming that speculative execution was squashed before gadgets could execute.

\vspace{-0.8em}
\section{Janus: Design and Implementation}

\begin{figure}[!b]
    \centering
        \captionsetup{skip=0pt}
        \includegraphics[width=0.75\linewidth]{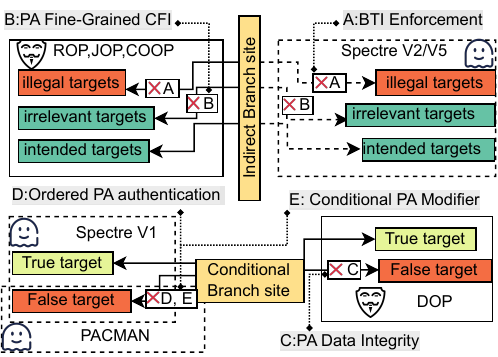}
        \caption{Janus Defense Mechanisms}
        \label{fig:janus_mechanisms} %
        \vspace{-0.5em}
\end{figure}

\subsection{Core Mechanisms}

Janus defends against three key vulnerabilities: CFI hijacking, Data-Oriented Programming (DOP)\cite{hu2016dop}, and speculative execution attacks. \cref{fig:janus_mechanisms} illustrates how the five core mechanisms of Janus (A–E) address these threats. 

A. The BTI enforcement targets indirect branches with BTI instructions, ensuring speculative execution only targets legitimate locations and preventing control-flow hijacking, such as ROP/JOP.

B. The PA Fine-Grained CFI uses Arm PA to assign a unique \texttt{modifier} to each function pointer, verifying that indirect branch targets transfer control flow to the intended destination. 

C. The PA Data Integrity signs critical DOP variables, ensuring the validity of data sources used for branch decisions and mitigating DOP attacks.

D. The Ordered PA Authentication mechanism enforces a strict sequence, ensuring that PA instructions precede DOP variable verification. This prevents attackers from reading the PA signature that follows, thwarting PACMAN attacks. 

E. The Conditional PA Modifier mechanism encodes the true outcome of conditional branches into a PA \texttt{modifier}. If the branch mispredicts, the PA verification fails and aborts speculative execution, defending against Spectre V1 attacks.

Through the synergistic interaction of these mechanisms, Janus establishes a comprehensive and efficient framework for defending against speculative execution attacks.

\begin{figure*}[!t]
    \centering %

    \begin{minipage}[t]{0.28\textwidth}
        \centering %
        \includegraphics[width=0.92\linewidth]{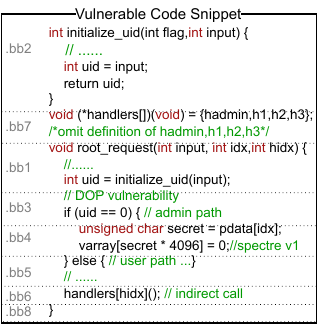} 
        \captionof{figure}{Vulnerable C Code Snippet.} %
        \label{fig:poc_unprotected_cpp}
    \end{minipage}
    \hfill %
    \begin{minipage}[t]{0.35\textwidth}
        \centering
        \includegraphics[width=0.94\linewidth]{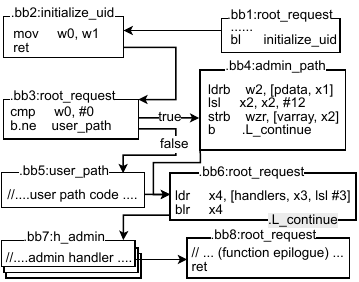}
        \captionof{figure}{CFG of Unprotected Assembly Code.} %
        \label{fig:poc_unprotected_asm}
    \end{minipage}
    \hfill
    \begin{minipage}[t]{0.35\textwidth}
        \centering
        \includegraphics[width=0.94\linewidth]{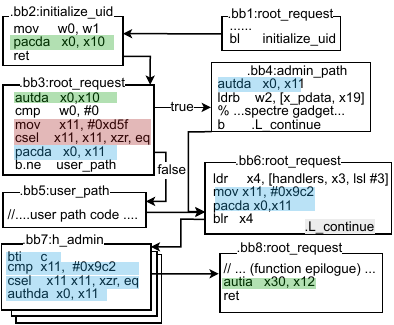}
        \captionof{figure}{CFG of Janus-Hardened Assembly Code.} %
        \label{fig:poc_janus_asm}
    \end{minipage}

    \caption*{\footnotesize{*Janus's Instrumentation for Unified Architectural and Microarchitectural Defense. Instrumentation for architectural Control Integrity is highlighted in \colorlegend{codegreen30}, mitigations for Spectre attacks in \colorlegend{codered30}, and dual-purpose instructions providing unified protection in \colorlegend{codeblue30}.}}
    \vspace{-0.5em}
\end{figure*}

\subsection{Janus Analysis and Instrumentation Pipeline}

Janus implements the core defense mechanisms as a multi-pass LLVM compiler framework, as shown in \cref{fig:janus_implement}. The pipeline starts at the IR level, where analysis passes identify control-flow structures. The \texttt{GetBranchTarget} and \texttt{GetBranchInfo} passes handle CFI and DOP analysis, respectively. CFI analysis identifies indirect branch call sites and their potential targets, creating an equivalence class for each site, while DOP analysis flags conditional branches influenced by external inputs as requiring protection.

The Janus defense framework is independent of these analysis techniques and focuses on efficient enforcement, allowing the replacement of current analysis with more precise static or dynamic methods, such as multi-layer type analysis (MLTA) for CFI or inter-procedural taint analysis for DFI.

The \texttt{AnalysisBTITarget} and \texttt{AnalysisPATarget} passes use mappings from internal or external analyses to generate PA modifiers for injection and verification. Janus’s policy-agnostic design ensures consistent PA-based operations, supporting various CFI/DFI policies without altering the core enforcement code. The metadata required for instrumentation is passed securely to the MachineIR level through a custom global structure, \texttt{IRsaveInfo}, simplifying inter-pass communication. Finally, passes like \texttt{InstruBTIMIR} and \texttt{InstruPAMIR} inject BTI and PA instructions into Machine Basic Blocks (MBBs), ensuring both precision and efficiency in the final code generation.

\subsection{From Vulnerability to Hardening: The Janus Workflow}

\cref{fig:poc_unprotected_cpp} illustrates a C code snippet with a DOP vulnerability, Spectre V1 vulnerability, and an indirect call vulnerability, its unprotected Control-Flow Graph (CFG)(\cref{fig:poc_unprotected_asm}), and the hardened assembly fragments after Janus protection (\cref{fig:poc_janus_asm}). These figures will help clarify how Janus:

\begin{itemize}
    \item Applies existing CFI/DFI analysis results.
    \item Extends CFI/DFI defenses to protect against control-flow Spectre attacks.
\end{itemize}

\subsubsection{Applying CFI/DFI Analysis to Establish Architectural Defenses}

As a policy-agnostic enforcement platform, Janus can translate external static or dynamic security policies into hardware-enforced protections for both CFI and DFI.

For DFI, Janus uses the ``Source-to-Sink'' model. It reads external taint analysis policies and applies the taint tag IDs as PA modifiers. Instrumentation is performed at the specified ``source'' (definition points) and ``sink'' (use points). For example, as shown in basic blocks \texttt{.bb2} and \texttt{.bb3}, Janus inserts a \texttt{pacda} instruction to sign the data at its generation point and an \texttt{autda} to verify it at its use point, ensuring data-flow integrity.

For CFI, Janus employs a hybrid scheme. Unlike traditional hardware CFI, which verifies directly at the call site, Janus avoids tracking the \texttt{PAC} lifecycle. Compared to software CFI schemes, Janus cryptographically hardens the checkpoint, addressing microarchitectural weaknesses. As shown in basic block \texttt{.bb6}, Janus inserts a lightweight \texttt{mov x11, \#0x9c2} and \texttt{pacda x0, x11} at the indirect call site to encode the control-flow intent. At the callee entry (\texttt{.bb7}), a BTI instruction ensures verification. The \texttt{cmp} and \texttt{csel} instructions compare the incoming tag, and if the comparison fails, \texttt{csel} clears \texttt{x11}, causing the subsequent \texttt{authda x0, x11} verification to fail, trapping control-flow hijacking.

\subsubsection{Extending Foundational Defenses to Mitigate Spectre Attacks}

Janus reuses its architectural control-flow integrity mechanisms to defend against speculative execution attacks targeting control flow.
For conditional branch mispredictions (Spectre V1), Janus uses the PA mechanism from its DFI protection. As shown in \texttt{.bb3}, the \texttt{csel} and \texttt{pacia} instructions dynamically encode the branch’s true outcome into the PA modifier (\texttt{x11}). During misprediction, the speculative path carries an incorrect modifier, causing the \texttt{autia} verification in \texttt{.bb4} to fail and squashing the pipeline, preventing Spectre gadgets from leaking data. This approach also defends against side-channel attacks like PACMAN, as any attempt to leak a PAC on a mis-speculated path is thwarted by the premature \texttt{aut*} verification failure.

For indirect branch mispredictions (Spectre V2), Janus introduces a ``cryptographic assertion at entry'' mechanism. As shown in \texttt{.bb7}, the \texttt{cmp} and \texttt{csel} sequence at the function entry creates a speculation barrier. If a Spectre-V2 attack mispredicts the entry, the incorrect context causes the comparison \texttt{cmp x11, \#0x9c2} to fail, and \texttt{csel} selects an invalid sentinel value (e.g., \texttt{xzr}). The following \texttt{authda} verification fails, squashing speculation before any potentially dangerous code executes.

This mechanism also protects against Spectre V5 attacks caused by RSB mispredictions. A mispredicted return that jumps to \texttt{.bb7} triggers the same verification failure, preventing execution.

\subsection{Cross-Domain Security Policy Fusion and Performance Optimization}

The core advantage of Janus lies in its comprehensive security coverage and synergistic optimization framework. It fuses policies from different analyses (CFI, DFI, Spectre) and reuses hardware-level instructions, providing robust protection with minimal overhead.
Janus employs two key optimization mechanisms to reduce overhead:

Modifier Fusion (MF): Traditional defense approaches insert isolated protection instructions for different threats (e.g., data-flow hijacking and side-channel attacks), leading to redundant performance costs. For example, a variable protected by both DFI and Spectre policies triggers two separate PA signing operations. To address this, Janus uses a Cross-Domain Policy Fusion mechanism, identifying variables requiring dual protection, combining their PA modifiers (\texttt{mod\_DFI} and \texttt{mod\_Spectre}) using XOR, and instrumenting a single PA instruction with the fused modifier (\cref{alg:multi_threat_pa_modifier_fusion}). This reduces instruction overhead, minimizes tag collisions, and simplifies policy management.

\begin{algorithm}[htbp]
  \caption{Multi-Threat PA Modifier Fusion}
  \label{alg:multi_threat_pa_modifier_fusion}
  
  \KwIn{
    $S_{DFI}, S_{Spectre}$: Sets of policies as $(v, loc, mod)$ tuples.
  }
  \KwOut{
    $I_{Optimized}$: Optimized set of instrumentation instructions.
  }
  
  \BlankLine
  
  Initialize $PolicyMap \leftarrow \emptyset$\;
  Initialize $I_{Optimized} \leftarrow \emptyset$\;
  
  \ForEach{policy $(v, loc, mod)$ in $S_{DFI} \cup S_{Spectre}$}{
    $PolicyMap[(v, loc)] \leftarrow PolicyMap[(v, loc)] \cup \{\text{policy}\}$\;
  }
  
  \BlankLine
  
  \ForEach{key $(v, loc)$ in $PolicyMap.\texttt{keys}()$}{
    $policies \leftarrow PolicyMap[(v, loc)]$\;
    
    $mod_{DFI} \leftarrow \texttt{get\_modifier\_or\_default}(policies, \text{DFI}, 0)$\;
    $mod_{Spectre} \leftarrow \texttt{get\_modifier\_or\_default}(policies, \text{Spectre}, 0)$\;
    
    $mod_{Fused} \leftarrow mod_{DFI} \oplus mod_{Spectre}$\;
    
    $instr \leftarrow \texttt{GeneratePA}(v, loc, mod_{Fused})$\;
    $I_{Optimized}.\texttt{add}(instr)$\;
  }
  
  \BlankLine
  \KwRet{$I_{Optimized}$}\;
\end{algorithm} %

Carrier Reuse (CR): Unlike traditional CFI schemes, Janus reduces overhead by reusing existing registers for CFI verification. Instead of allocating a new register, it searches for an existing variable (e.g., DFI-protected) to carry the CFI context. As shown in \cref{alg:fusing_cfi_context}, Janus "piggybacks" the CFI verification onto the existing PA verification, merging DFI and CFI checks into a single operation with zero additional register overhead. If no suitable carrier is found, it checks if a stack-related register, such as \texttt{LR}, can be used; otherwise, it falls back to a general-purpose register like the non-volatile register \texttt{x10}.

\begin{algorithm}[htbp]
  \caption{Fusing CFI Context into Existing PA Policies}
  \label{alg:fusing_cfi_context}
  
  \SetKwProg{Fn}{Procedure}{}{}

  \KwIn{
    $S_{CFI}$: Indirect call sites as $(call\_site, cfi\_tag)$ tuples. \\
    $P_{Existing}$: Existing PA policies as $(variable, loc, mod)$ tuples.
  }
  \KwOut{
    $I_{Final}$: Final set of instrumentation instructions.
  }

  Initialize $I_{Final} \leftarrow \emptyset$\;
  
  \BlankLine
  
  \ForEach{$(call\_site, cfi\_tag)$ in $S_{CFI}$}{
    $tag\_reg \leftarrow \texttt{GetTagPassingRegister}()$\; %
    $carrier, orig\_mod \leftarrow \texttt{FindCarrier}(call\_site, P_{Existing})$\;
    $callee \leftarrow \texttt{GetCalleeFrom}(call\_site)$\;
    $mod\_reg \leftarrow \texttt{AllocateScratch}()$\;
    $I_{Final}.\texttt{add}(\texttt{CreateInst}(\text{"MOV"}, tag\_reg, cfi\_tag), \text{ at caller epilogue})$\;
    $I_{Final}.\texttt{add}(\texttt{CreateInst}(\text{"CMP"}, tag\_reg, cfi\_tag), \text{ at callee})$\;

    \uIf{$carrier$ is not \textbf{null}}{
      $I_{Final}.\texttt{add}(\texttt{CreateInst}(\text{"CSEL"}, mod\_reg, orig\_mod, \text{XZR, EQ}), \text{ at callee})$\;
      \vspace{-0.3cm}
      $I_{Final}.\texttt{add}(\texttt{CreateInst}(\text{"AUTHDA"}, carrier, mod\_reg), \text{ at callee})$\;
    }
    \ElseIf{\texttt{IsNonLeaf}(callee)}{
      $I_{Final}.\texttt{add}(\texttt{CreateInst}(\text{"CSEL"}, mod\_reg, \text{SP, XZR, EQ}), \text{ at callee})$\;
      $I_{Final}.\texttt{add}(\texttt{CreateInst}(\text{"AUTHIA"}, \text{LR}, mod\_reg), \text{ at callee epilogue})$\;
    }
    \BlankLine
    $I_{Final}.\texttt{add}(\texttt{CreateInst}(\text{"CMP"}, tag\_reg, \texttt{GetTagOf}(callee)), \text{ at callee entry})$\;
  }
  
  \BlankLine
  \KwRet{$I_{Final}$}\;
\end{algorithm}

Finally, Janus optimizes performance through both compiler-level and instruction-level techniques, including LLVM’s Link-Time Optimization (LTO), function inlining, and loop-aware optimizations to eliminate redundant pointer re-signing in safe loop bodies.

\section{Implementation}

We implemented a Janus prototype as out-of-tree passes for the LLVM compiler (v20.1.0). The framework has three parts: 

\textbf{Basic CFI and Naive DOP-sensitive variable detection:} Implements simple type-based CFI and basic DOP-sensitive variable detection, including branch variables and taint analysis for a single call level. Janus supports more complex CFI/DFI schemes, but for prototyping transient execution defenses, we kept the CFI/DFI strength simple. This part comprises 3,700 C++ SLOC.
\textbf{External analysis results reader:} A 600-line program that reads external analysis results.
\textbf{Instrumentation:} Inserts necessary instructions at the \texttt{MachineIR} level. 

Instrumentation occurs after optimization but before register allocation, ensuring security gadgets remain unaffected by later optimizations and preserving \texttt{PA modifier} register availability. The out-of-tree design enhances portability, and a binary validator checks the disassembled machine code against IR-level metadata to confirm protection correctness.

\begin{figure}[hbt]
    \centering %
    \includegraphics[width=0.92\columnwidth]{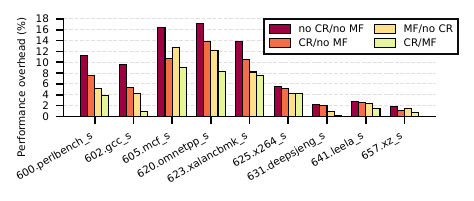}
    \caption{SPEC CPU 2017 performance results.}
    \label{fig:janus_perf}
\vspace{-0.5em}
\end{figure}

\section{Evaluation}
\label{sec:7:evaluation}
\subsection{Experimental Setup}
\label{subsec:setup}

We use a unified strategy for performance and security evaluation on the \texttt{Radxa Orion O6} development board with 16 GB of \texttt{DDR5} memory, featuring an \texttt{ARM Cortex-A720} CPU with 4 big cores at 2.8 GHz, 4 medium cores at 2.4 GHz, and 4 little cores at 1.8 GHz, running \texttt{Ubuntu 24} (\texttt{Kernel 6.11}) with \texttt{LLVM/Clang 20.1} (+\texttt{LTO}, -\texttt{O3}).
For consistency, we fixed the CPU to the 4 big \texttt{A720} cores at 2.8 GHz. The platform provides direct hardware access for running PoC exploits and measuring microarchitectural phenomena via performance counters, eliminating thermal throttling and virtualization-induced jitter, ensuring accurate SPEC CPU 2017 benchmarking and security validation.
Experiments were conducted in a minimal, single-user Linux environment with CPU frequencies locked to maximum and benchmark processes pinned to dedicated cores using \texttt{taskset}.

\subsection{Performance Evaluation}

\subsubsection{SPEC CPU 2017}

We evaluate Janus’s runtime overhead using the SPEC CPU 2017 benchmark suite, specifically the SPEC speed Integer suite, which includes nine C/C++ applications. Using the uninstrumented build as the baseline, we compare performance with \texttt{modifier fusion (MF)}, \texttt{carrier reuse (CR)}, and all optimizations (including \texttt{LTO}) enabled, as shown in Figure 7.

We report the geometric mean of overhead across five runs. The unoptimized version has an overhead of 8.17\%. With \texttt{MF}, it drops to 5.74\%; with \texttt{CR}, it further drops to 5.27\%; and with all optimizations, it reduces to 3.85\%. In most tests, \texttt{MF} provides greater performance improvement, while \texttt{CR} has a more significant effect in tests like 605 and 657, likely due to higher register pressure where \texttt{CR}’s register optimization is more beneficial. The above represents the overall overhead of the framework. Additionally, we measured the net overhead specifically for instructions used in speculative execution defense. By subtracting the overhead of strip-janus (which replaces and removes Spectre-related protection instructions such as csel) from the overhead of full-janus, the average overhead for the SPEC CPU 2017 test is 0.58\%.

\subsubsection{Real-World Applications}

We evaluated the performance overhead on two real-world applications: Nginx (I/O-intensive) and SQLite (CPU/memory-intensive).

\textbf{Nginx}: We benchmarked Nginx using the HTTP tool \texttt{wrk}, generating HTTP requests for one minute with 4 threads and 128 concurrent connections. The test was run three times, serving 1KB, 100KB, and 1MB files, with 4, 4, and 2 worker threads to maximize CPU utilization. The throughput degradation ranged from 0.13\% to 2.97\%.

\textbf{SQLite}: We stress-tested SQLite using its \texttt{Speedtest} benchmark, which executes database operations and reports total execution time. The test used an in-memory database to isolate CPU overhead, with default settings and compiler annotations to avoid unnecessary instrumentation. With the added protection, the average completion time for SQLite increased by 7.80\%.

\begin{table}[hb]
\centering
\caption{Asymmetrical ablation study of Janus overhead on SQLite and Nginx.}
\label{tab:janus_asymmetrical}

\footnotesize
\renewcommand{\arraystretch}{1.0}
\setlength{\tabcolsep}{3pt} %

\begin{tabular}{llrrrr}
\toprule

\textbf{Benchmark} & \textbf{Config.} & \multicolumn{3}{c}{\textbf{Perf. Ov. (\%)}} & \textbf{Code Size Ov. (\%)} \\
\midrule

\multirow{4}{*}{\textbf{SQLite}}
 & w/o opts    & \multicolumn{3}{c}{9.58}  & 16.57 \\
 & w/ CR only  & \multicolumn{3}{c}{9.15}  & 9.86 \\
 & w/ MF only  & \multicolumn{3}{c}{8.27}  & 8.24 \\
 & \textbf{Full}     & \multicolumn{3}{c}{\textbf{7.80}} & \textbf{8.32} \\
\midrule

\multirow{6}{*}{\textbf{Nginx}}
 & & \textbf{1k} & \textbf{100k} & \textbf{1m} & \\
 \cmidrule(lr){3-5}
 & w/o opts    & 0.95 & 3.85 & 1.34 & 19.84 \\
 & w/ CR only  & 0.83 & 3.61 & 1.07 & 16.61 \\
 & w/ MF only  & 0.32 & 3.23 & 0.82 & 15.14 \\
 & \textbf{Full}     & \textbf{0.13} & \textbf{2.97} & \textbf{0.48} & \textbf{13.42} \\
\bottomrule
\end{tabular}
\end{table}

\subsubsection{Code Size Increase Evaluation}

We measured Janus’s impact on binary code size for the SPEC CPU 2017 benchmarks and two real-world applications. As shown in Figure 8, the code size overhead for Janus ranges from 8.32\% to 13.42\% for the real-world applications and 2.8\% to 20.2\% for the SPEC CPU 2017 benchmarks.

\begin{figure}[thb]
    \centering %
    \includegraphics[width=0.92\columnwidth]{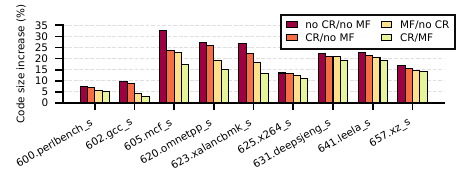}
    \caption{Code-size increase for Janus.}
    \label{fig:janus_size}
\vspace{-0.5em}
\end{figure}

\subsection{Security Evaluation}

\textbf{CFI and DFI:} Janus defends against Spectre attacks by leveraging external CFI/DFI analysis results and using basic non-control data protection (conditional variables) for testing. It passed all ConFIRM tests except JIT and effectively protected against string buffer attacks on conditional branch variables.

\textbf{Traditional Spectre Attacks:} To evaluate Janus’s effectiveness against transient execution attacks, we implemented PoC attacks for \texttt{Spectre V1}, \texttt{V2}, and \texttt{V5} on the ARM platform, using the \texttt{Radxa Orion O6} development board. The baseline PoC attacks successfully leaked secrets via a cache side channel. After applying Janus, all attacks were mitigated, with no secrets recovered, and cache probes resulted in misses, confirming speculative execution paths were squashed before accessing sensitive data.

The key advantage of Janus’s defense is its insensitivity to diverse leakage gadget patterns. Unlike defenses that rely on gadget identification, Janus focuses on ensuring the legitimacy of control flow in speculative execution. It reuses the microarchitectural side effects of PA and BTI to enforce fine-grained verification of speculative correctness at the branch instruction, the chokepoint for all control-flow attacks.
We validated Janus by constructing various leakage gadgets, showing that the defense remained effective regardless of the gadget's structure. The attack was intercepted at its source, preventing it from reaching the gadget. This demonstrates that Janus offers fundamental, forward-looking protection rather than relying on pattern matching.

\textbf{PACMAN:} We also tested Janus’s mitigation of PACMAN attacks, which aim to leak PA metadata. Using the original PoC exploit, we adapted it for our testbed and confirmed its ability to steal a \texttt{PAC}. After applying Janus protection, the exploit failed to steal the \texttt{PAC}, with no cache signal detected.

\section{Related Work}

Janus draws on the SLH-enhanced method \cite{christou2024eclipse,vassena2021automatically,marinaro2024beyond}  to control speculative execution and enhance protection using memory safety hardware. While FineIBT reduces the speculation window, it doesn’t fully eliminate it. Existing ARM-based Spectre defenses like SpecASan \cite{specasan} (hardware) and SwitchPotline \cite{bauer2024switchpoline} (software) target specific vulnerabilities. In contrast, Janus provides a comprehensive defense against control-flow related speculative execution attacks, integrating CFI and memory safety mechanisms for stronger, optimized protection.

\section{Conclusion}

In this paper, we introduced Janus, a security framework that mitigates transient execution attacks using hardware memory safety primitives. By combining Arm PA and BTI, Janus ensures control-flow and data integrity, defending against Spectre variants and PACMAN attacks. Evaluation on \texttt{ARMv8-A} hardware shows Janus provides robust protection with 3.85\% average overhead, demonstrating its effectiveness and cross-platform adaptability. Janu offers a scalable, efficient solution for safeguarding against speculative execution vulnerabilities.

\begin{acks}
This work was supported in part by the National Key Research and Development Program of China under Grant No.~2024YFE0211100, the Joint Funds of the National Natural Science Foundation of China under Grant No.~U24A6009, and the National Science Fund for Distinguished Young Scholars under Grant No.~62125208.
\end{acks}

\clearpage
\bibliographystyle{bib/ACM-Reference-Format}
\bibliography{bib/reference}

\end{document}